\documentclass[12pt]{article}

\usepackage{amsmath, amsfonts,amssymb,amsthm,titling,blindtext, latexsym,epsfig,color,rotating,paralist,times,float,subfigure,algorithm,verbatim}

\usepackage{geometry,setspace,times,ifthen}
\usepackage{fancyhdr}
\usepackage{tikz}

\makeatletter
\def\nl#1#2{\begingroup
    #2%
    \def\@currentlabel{#2}%
    \phantomsection\label{#1}\endgroup
}
\makeatother

{\ensuremath{
\begin{empheq}[box=\fbox]{align}
{#1}
\end{empheq}
}}

{
\begin{mdframed}
\par\noindent\textbf{#1:}\begin{rmfamily}\noindent}%
{\end{rmfamily}
\end{mdframed}
}

\newcommand{\sgn}{\operatorname{sgn}}

\newtheorem{theorem}            {Theorem}[section]

\newtheorem{definition}         [theorem]{Definition}

\newtheorem{lemma}              [theorem]{Lemma}

\DeclareMathOperator*{\argmax}{arg\,max}

\newcommand{\pdf}{p}

\newcommand{\prob}{\mathbb{P}}
\newcommand{\E}                 {\Bbb{E}}

\newcommand{\incv}{f}

\newcommand{\obs}{y}

\newcommand{\snoise}{w}

\newcommand{\state}{x}
\newcommand{\statespace}{\mathcal{X}}
\newcommand{\obspace}{\mathcal{Y}}
\newcommand{\statedim}{X}
\newcommand{\obsdim}{{Y}}

\newcommand{\fun}{\phi}

\newcommand{\oprob}{B}
\newcommand{\tp}{P}

\newcommand{\model}{\theta}

\newcommand{\belief}{\pi}

\newcommand{\bbelief}{\bar{\pi}}

\newcommand{\Belief}{\Pi(\statedim)}

\newcommand{\tpi}{\pi_2}

\newcommand{\ole}{\stackrel{\text{defn}}{=}}

\newcommand{\gr}{\geq_r}
\newcommand{\lr}{\leq_r}
\newcommand{\gs}{\geq_s}
\newcommand{\ls}{\leq_s}

\newcommand{\filterd}{\sigma}

\newcommand{\filter}{T}

\newcommand{\reals}{{\rm I\hspace{-.07cm}R}}

\newcommand{\beq}{\begin{equation}}
\newcommand{\eeq}{\end{equation}}
\newcommand{\nn}{\nonumber}

\newcommand{\p}{\prime}

\newcommand{\one}{\mathbf{1}}
\newcommand{\ones}{\mathbf{1}}

\newcommand{\diag}{\textnormal{diag}}

\newcommand{\lR}{\preceq}

\newcommand{\bmodel}{\bar{\model}}

\newcommand{\reward}{r}

\newcommand{\action}{u}

\newcommand{\actionspace}{\,\mathcal{U}}
\newcommand{\actiondim}{U}

\newcommand{\discount}{\rho}

 \newcommand{\Ep}{\E_{\policy}}

\newcommand{\policy}{\mu}

\newcommand{\optpolicy}{\policy^*}

\newcommand{\valuef}{V}

\renewcommand{\time}{k}

\newcommand{\Hyperplane}{\mathcal{H}}

\renewcommand{\l}{\mathcal{L}}

\newcommand{\copomat}{\Gamma}

\newcommand {\policyl} {\underline{\mu}}

\newcommand{\bj}{l}

\newcommand{\gl}{\geq_{L_i}}

\newcommand{\glX}{\geq_{L_X}}
\newcommand{\glone}{\geq_{L_1}}

\newcommand{\bp}{{\bar{\pi}}}

\newcommand{\barray}{\begin{array}{ll}}
\newcommand{\earray}{\end{array}}

\title{New Sufficient Conditions for Lower Bounding the Optimal Policy of a  POMDP using Lehmann Precision}

\author{Vikram Krishnamurthy,\\
 Electrical \& Computer
Engineering, \\ Cornell Tech \\ Cornell University, USA. \\
vikramk@cornell.edu}

\begin{document}
\begin{titlingpage}\maketitle
  
\begin{abstract}
  This paper provides new sufficient conditions  so that the optimal policy of a
  partially observed Markov decision process (POMDP) can be lower bounded by a myopic policy. The two new proposed conditions, namely, Lehmann precision and copositive dominance, completely fix the problems with two crucial assumptions in the well known papers \cite{Lov87,Rie91}. For controlled sensing POMDPs, Lehmann precision
exploits both convexity and monotonicity of the value function, whereas the classical Blackwell dominance only exploits convexity.
Numerical examples  are presented where   Lehmann precision holds but  Blackwell dominance does not hold, thereby illustrating the usefulness of the main result in controlled sensing applications.
\end{abstract}
\end{titlingpage}

\section{Introduction} \label{sec:intro}
This paper provides  sufficient conditions so that  the optimal policy of a POMDP is provably  lower bounded by a myopic policy. From a practical point of view, this  structural result is useful since  myopic policies are trivial to compute/implement in large scale POMDPs  and also provide  a useful initialization for more sophisticated sub-optimal solutions.
Structural results  are important since in general solving  POMDPs  is  PSPACE-complete; see
\cite{PT87}.

The seminal papers \cite{Lov87,Rie91,RZ94} give sufficient conditions for two very useful results: (i) the value function of a POMDP to be  monotone in the belief state (with respect to the likelihood ratio order and  multivariate generalizations) and (ii)  for the optimal policy of a POMDP to be lower bounded by a myopic policy. Monotonicity  of the value function is  crucially important and will be used in our main results below. Regarding lower bounding the optimal policy by a myopic policy, unfortunately, despite the  enormous usefulness of such
a result, the sufficient conditions given in \cite{Lov87} and \cite{Rie91} are not useful - it is impossible to generate non-trivial examples that satisfy the conditions (c), (e), (f) of \cite[Proposition 2]{Lov87} and condition (i) of \cite[Theorem 5.6]{Rie91}.
Our recent works \cite{Kri16,KP15} provided a fix for the  conditions on the transition probabilities  by using copositive dominance.
In this paper, motivated by controlled sensing applications, we provide a complete fix to the conditions on the controlled observation probabilities of the POMDP so that   the results of \cite{Lov87,Rie91} hold for constructing a myopic policy that lower  bounds the optimal policy.

This paper is motivated by controlled sensing POMDPs  where the observation probabilities (which model an adaptive sensor) are controlled whereas  the transition probabilities  (which model the Markov chain signal being observed by the sensor) are not controlled.
Controlled sensing arises in a variety of applications in reconfigurable sensing (how can a sensor reconfigure its behavior in real time), cognitive radio, adaptive radars, optimal search problems for a Markovian target, and active hypothesis  testing.
Providing    useful sufficient conditions so that the optimal policy for a controlled sensing POMDPs is lower bounded by a myopic policy is surprisingly  nontrivial. The main new assumption that we will use  is the {\em Lehmann precision} condition    --  this single crossing 
condition proposed in \cite{Miz06} has recently been used extensively in the economics  literature, see \cite{GP10,AL18}. Thus far,  there has been no way of obtaining structural results for  controlled sensing POMDPs that exploit both
monotonicity {\em and} convexity of the value function. The papers
 \cite{Lov87,Rie91} used only monotonicity of the value function (wrt monotone likelihood ratio stochastic order) and the  resulting assumptions  were not useful (as mentioned above).  On the other hand,  \cite{WD80,Rie91,RZ94} 
used only convexity of the value function with  Blackwell dominance
to construct a lower bound to a controlled sensing POMDP.  In this paper,
Lehmann precision allows us to use both convexity and monotonicity of the value function to construct the lower  bound. Indeed, the Lehmann precision  condition on the observation probabilities   together with copositive dominance of controlled transition matrices, gives a useful set of conditions for POMDPs which completely fix the problems with the key assumptions in  \cite{Lov87} and also
\cite{Rie91}.  Theorem \ref{thm:main} is our main POMDP structural result.

In proving our main result, as an aside we also establish two minor results.
First, Theorem \ref{thm:ordinal} compares the optimal cumulative rewards of two different POMDPs when the parameters of one dominate the other with respect to Lehmann precision; the result is more useful
 than the Blackwell dominance case in controlled sensing POMDPs.  Second, Theorem \ref{thm:valuedec}   cleans up the  assumption made in \cite{Alb79} which results in  the piecewise linear segments of the POMDP value function being monotone vectors. The assumption in \cite{Alb79} is  implicit and not easily verifiable.  Our proof uses stochastic dominance restricted to certain line segments to show that   the conditions in  \cite{Lov87} actually do result in monotone vectors for the value function for the case of 3 or fewer underlying states.

\section {The Partially Observed Markov Decision Process} \label{sec:pomdp}

Consider  a discrete time, infinite horizon discounted reward POMDP. A   discrete time Markov chain  evolves on the  state space $\statespace = \{1,2,\ldots, \statedim\}$. Denote the
action space  as $\actionspace = \{1,2,\ldots,\actiondim\}$ and observation space as $\obspace$. We consider either  $\obspace =  \{1,2,\ldots,\obsdim\}$ (finite set) or $\obspace =  \reals$  or $\obspace$ is the closed interval $[1,\obsdim]$.
Let
$\Belief = \left\{\belief: \belief(i) \in [0,1], \sum_{i=1}^\statedim \belief(i) = 1 \right\}$ denote the belief space of $\statedim$-dimensional probability vectors.  For stationary policy  $\policy: \Belief \rightarrow \actionspace$,
 initial belief  $\belief_0\in \Belief$,  discount factor $\discount \in [0,1)$, define the  discounted cumulative reward:
\begin{align}\label{eq:discountedcost}
J_{\policy}(\belief_0) = \Ep\left\{\sum_{\time=0}^{\infty} \discount ^{\time}\, \reward_{\policy(\belief_\time)}^\p\, \belief_\time\right\}.
\end{align}
Here $\reward_\action = [\reward(1,\action),\ldots,\reward(\statedim,\action)]^\p$, $u\in \actionspace$ is the reward vector for each sensing action, and the belief state evolves according to Bayes formula as
$\belief_{k} = \filter(\belief_{k-1},\obs_k,\action_k)$ where
\begin{align}  \filter\left(\belief,\obs,\action\right) = \cfrac{\oprob_{\obs} (\action)\, \tp^\p(\action) \belief}{\filterd\left(\belief,\obs,\action\right)} , \quad
\filterd\left(\belief,\obs,\action\right) = \one_{\statedim}'\oprob_{\obs}(\action) \tp^\p(\action)\belief, \quad
\oprob_{\obs}(\action) = \diag\{\oprob_{1,\obs}(\action),\cdots,\oprob_{\statedim,\obs}(\action)\}. \label{eq:information_state}
\end{align}
Here  $\one_{\statedim}$ represents a $\statedim$-dimensional vector of ones,
$ \tp(\action) = \left[\tp_{ij}\right]_{\statedim\times\statedim}$
$ \tp_{ij}(\action) = \prob(\state_{\time+1} = j | \state_\time = i,\action_k = \action )$ denote the controlled  transition probabilities. When $\obspace$ is a finite set,
$\oprob_{\state\obs}(\action) = \prob(\obs_{\time+1} = \obs| \state_{\time+1} = \state, \action_{\time} = \action)$ denotes the controlled observation   probabilities; for $\obspace$ continuum, we assume 
that the conditional distribution
$\prob(\obs_k \leq y| \state_k)$ is absolutely continuous wrt the Lebesgue measure and so the controlled conditional probability density function
$\oprob_{\state\obs}(\action) = \pdf(\obs_{\time+1} = \obs| \state_{\time+1} = \state, \action_{\time} = \action)$ exists.

The aim is to compute the optimal  stationary policy $\optpolicy:\Belief \rightarrow \actionspace$ such that
$J_{\optpolicy}(\belief_0) \leq J_{\policy}(\belief_0)$ for all $\belief_0 \in \Belief$.
Obtaining the optimal policy  $\optpolicy$ is equivalent to solving
 Bellman's  dynamic programming equation:
$ \optpolicy(\belief) =  \underset{\action \in \actionspace}\argmax~ Q(\belief,\action)$, $J_{\optpolicy}(\belief_0) = \valuef(\belief_0)$, where
\begin{equation}
\valuef(\belief)  = \underset{\action \in \actionspace}\max ~Q(\belief,\action), \quad
  Q(\belief,\action) =  ~\reward_\action^\prime\belief + \discount\sum_{\obs \in \obsdim} \valuef\left(\filter\left(\belief,\obs,\action\right)\right)\filterd \left(\belief,\obs,\action\right). \label{eq:bellman}
\end{equation}
Note that for continuum $\obspace$, the notation $\sum_{\obs \in \obspace}$ denotes integration wrt $y$.
Also,  $\valuef(\belief)$ is the fixed point of the following value iteration algorithm: Initialize $V_0(\belief)=0$ for $\belief \in \Belief$. Then
\begin{equation}
  \begin{split}
  \valuef_{k+1}(\belief)  &= \underset{\action \in \actionspace}\max ~Q_{k+1}(\belief,\action), \quad \mu_k = \argmax_{\action \in \actionspace} Q_k(\belief,\action), \\
  Q_{k+1}(\belief,\action) &=  ~\reward_\action^\prime\belief + \discount\sum_{\obs \in \obsdim} \valuef_k\left(\filter\left(\belief,\obs,\action\right)\right)\filterd \left(\belief,\obs,\action\right),\quad k=0,1,\ldots, \end{split} \label{eq:vi}
\end{equation}
Indeed, the sequence $\{V_k(\belief), k=0,1,\ldots\}$ converges uniformly  to $V(\belief)$ on $\Belief$ geometrically fast.
Since  $\Belief$ is continuum, Bellman's equation \eqref{eq:bellman} and the value iteration algorithm (\ref{eq:vi}) do not directly translate into practical solution methodologies since they need to be evaluated at each $\belief \in \Belief$.
Almost 50 years ago,  \cite{Son71} showed that when $\obspace$ is finite, then for any $k$,
$\valuef_k(\belief) $ has a finite dimensional  piecewise  linear and convex characterization.  
Unfortunately, the number of piecewise linear segments can increase exponentially with the action space dimension
$\actiondim$ and double exponentially with time $k$.
Thus there is strong  motivation for structural results:  to  construct useful myopic lower  bounds   ${\policyl(\belief)}$ for the optimal policy $\optpolicy(\belief)$. 

{\em Remark. Controlled Sensing}: In controlled sensing, the aim is to dynamically decide which sensor (or sensing mode) $\action_k$ to choose at each time $k$ to optimize the objective  (\ref{eq:discountedcost}).
For such POMDPs, the transition matrix $\tp$, which characterizes the dynamics of the signal being sensed, is  functionally independent of the action $\action$. Only $\reward_\action$, which models the information acquisition reward of the sensor, and observation probabilities $\oprob(\action)$, which models the sensor's accuracy when it operates in mode $\action$, are action dependent.

\section{Main Structural Result}
Although our main motivation stems from controlled sensing (where only the reward and observation matrix are action dependent), we state our main result for general  POMDPs where
the reward,  transition and observation matrices are action dependent; so that the results provide a complete fix to the conditions in  \cite{Lov87,Rie91}.
In particular, Assumptions A\ref{copositive} and A\ref{obssc}, A\ref{obsinit}  below provide a complete fix to   the problems inherent in
 conditions (c) and (f)  of \cite{Lov87}.

\begin{definition}[Copositive Ordering of Transition Matrices \cite{Kri16}]  \label{def:lR}
Given transition matrices $\tp(\action)$ and $\tp(\action+1)$, we say that   \index{copositive matrix}
$ \tp(\action) \lR \tp(\action+1)  $
if the sequence of $\statedim \times \statedim$ matrices $\copomat^{j,\action}$, $j=1\,\ldots,\statedim-1$ are copositive, i.e., 
\begin{align}
\belief^\p  \copomat^{j,\action} \belief & \geq 0 ,  \quad \forall \belief \in \Belief, \quad \text{ for each } j,
\text{ where }   \label{eq:tpdominance} \\ & \hspace{-1.2cm}
 \copomat^{j,\action} = \cfrac{1}{2}\left[\gamma^{j,\action}_{mn} + \gamma^{j,\action}_{nm}\right]_{\statedim\times\statedim},
\;
\gamma^{j,\action}_{mn} =  \tp_{m,j}(\action)\tp_{n,j+1}(\action+1) - \tp_{m,j+1}(\action)\tp_{n,j}(\action+1) . \nn
\end{align}
\end{definition}

Our main assumptions are the following:
\begin{enumerate}[\bf{(A}1)]

\item\label{dec_cost} [Monotone reward]  $\reward(i,\action)$ is increasing\footnote{Throughout this paper, by increasing, we mean non-decreasing.} in $i$ for  each $ \action \in \actionspace$.
  
\item\label{TP2_tp} [TP2 transition] $\tp(\action)$   is  totally positive of order 2 (TP2):  all second-order minors are nonnegative.\footnote{Equivalently, the $i$-th row is monotone likelihood ratio (MLR) dominated by the $(i+1)$-th row for $i=1,2,\ldots,\statedim-1$; MLR dominance is defined in Section~\ref{sec:proof}.}
  
\item\label{TP2_obs} [TP2 observation] $\oprob(\action)$,  $\action \in \actionspace$ is TP2.

\item \label{copositive} [Copositive dominance] $\tp(u)  \lR \tp(u+1)$

\item \label{obsdom} [Stochastic dominance of observations]
  $\sum_{\obs < j} \oprob_{i\obs}(\action) \leq \sum_{\obs < j} \oprob_{i\obs}(\action+1) $
  for all $i\in \statespace$ and $j \in \obspace$.  Equivalently,
  $\oprob_{i}(\action)  \ls  \oprob_{i}(\action+1)$ where
  $\oprob_i(\action)$ denotes the $i$-th row of observation matrix $\oprob(\action)$
  and $\ls$ denotes first order stochastic  dominance.
\item \label{obssc} [Lehmann precision]
  $\sum_{y\leq j} \oprob_{iy}(\action) - \sum_{y\leq \bj}  \oprob_{iy}(\action+1)$ changes sign at most once from negative to positive as $i$ increases for all $j,\bj \in \obspace$. We denote this as
  $\oprob(u+1) >_L \oprob(u)$.

\item \label{obsinit} If $\obspace = \reals$, then $\oprob_{iy}(\action+1)/ \oprob_{iy}(\action)< \infty$
  for $i=1,\ldots,\statedim$,
  i.e., absolute continuity holds.  \\ If  $\obspace= \{1,\ldots,\obspace\}$ (finite set)
  then  for the boundary values 1 and $\obspace$ and  $i=1,\ldots,\statedim$: \beq
  \oprob_{i1}(\action) \, \oprob_{\statedim 1}(\action+1)  \leq
  \oprob_{i1}(\action+1) \, \oprob_{\statedim 1}(\action),
  \quad
  \oprob_{i\obsdim}(\action) \, \oprob_{\statedim \obsdim}(\action+1)  \geq
  \oprob_{i\obsdim}(\action+1) \, \oprob_{\statedim \obsdim}(\action).
  \label{eq:obsinit}
  \eeq
  If $\obspace = [a,b]$ then (\ref{eq:obsinit}) holds with $1$ and $\obsdim$ replaced by $a$ and $b$.
  \end{enumerate}

  The single crossing  property A\ref{obssc} is called ``Lehmann precision'' in \cite{Jew07} and integral precision in \cite{GP10}; see also \cite{Leh88}.

  
  \begin{theorem}[Main Structural using Lehmann Precision] \label{thm:main}
    \begin{compactenum}
      \item Controlled Sensing POMDP: Suppose the transition  probabilities  $\tp$ are functionally independent  of the action, but the observation probabilities $\oprob(\action)$ are action dependent.  Assume    A\ref{TP2_tp},  A\ref{TP2_obs},   A\ref{obssc} (Lehmann precision),  A\ref{obsinit}  hold.
Then      $
     Q(\belief,\action) - \reward_\action^\p \belief  \uparrow \action $.
     Therefore, the myopic policy $\policyl(\belief)= \argmax_\action \reward_\action^\p \belief$  forms a lower bound to the optimal policy in the sense that
     $\optpolicy(\belief) \geq \policyl(\belief)$ for all $\belief \in \Belief$.
   \item General  POMDP: Suppose both the transition probabilities  $\tp(\action)$ and observation probabilities $\oprob(\action)$ are action dependent.
     Then  under A\ref{dec_cost},  A\ref{TP2_tp},  A\ref{TP2_obs},  A\ref{copositive} (copositive dominance), A\ref{obsdom}, A\ref{obssc} (Lehmann precision),  A\ref{obsinit}, the above result holds.
   \end{compactenum}
 \end{theorem}
 The proof is in Section \ref{sec:proof}. Theorem \ref{thm:main}  also holds for any finite horizon (with non-stationary policy).
 \subsection*{Discussion}
 From a practical point of view,   Theorem \ref{thm:main} is useful since  the myopic policy $\policyl$ is trivial to compute and implement and gives a guaranteed lower bound to the optimal policy.
Also, for beliefs $\belief$ where $\policyl(\belief)= \actiondim$, the optimal policy $\optpolicy(\belief)$ coincides with the myopic policy $\policyl(\belief)$.

The rest of this section discusses several implications of Theorem \ref{thm:main} and its assumptions.

1.  {\bf Assumptions}: Assumptions  A\ref{dec_cost} to A\ref{obsinit} along with Theorem \ref{thm:main} completely fixes the problems with the assumptions in 
 \cite{Lov87}  and  \cite{Rie91}. 

     Assumptions   A\ref{dec_cost},
     A\ref{TP2_tp}, A\ref{TP2_obs} and    A\ref{obsdom} correspond to conditions (a), (c), (d), (e) in  \cite[Proposition 1, Proposition  2]{Lov87}.
Indeed,  \cite{Lov87} proves that A\ref{dec_cost},
     A\ref{TP2_tp}, A\ref{TP2_obs} are sufficient for $\valuef(\belief)$ to increase with respect to $\belief$ (wrt monotone likelihood ratio order).

     \underline{(i) Assumption A\ref{dec_cost}}.  In Theorem \ref{thm:main},  A\ref{dec_cost}  (monotone rewards)
is only required for general POMDPs; it is not required for controlled sensing POMDPs.
Moreover,   for general POMDPs,  A\ref{dec_cost}  can be replaced by the following condition which depends only on the transition probabilities:
     \begin{enumerate}
       \item[{\bf (A1')}] 
         There exists $f \in \reals^\statedim$ such that $ \Delta_\action  \ole \big(I- \discount\,\tp(\action)\big) f $ 
is a strictly increasing vector for each action $u \in \actionspace.$
\end{enumerate}
  
 A1'  implies that there exists a POMDP with monotone increasing reward vectors $\reward_\action + \Delta_\action$ that has the same optimal policy as the original POMDP.
  To explain A1', suppose the reward vectors $r_u$, $u \in \actionspace$ are arbitrary;
not necessarily monotone. For   $\incv \in \reals^\statedim$, define $W(\belief) = V(\belief) + \incv^\p \belief$.
     Then it is easily seen that
$W(\belief)$ satisfies Bellman's equation (\ref{eq:bellman}) with 
reward vector $r_u + \Delta_u$, and the optimal policy remains unchanged.  Thus under A1' one can choose $\incv$ so that $r_u + \Delta_u$ is increasing, while the optimal policy remains unchanged.

For controlled sensing POMDPs, A1' always holds; hence Statement 1 of Theorem \ref{thm:main}  does not require  A\ref{dec_cost}.
Since  $\tp$ and $\Delta$ in A1'  do  not depend on $\action$, choose
 $\tilde{r}  > \max_{i,u,j,u'} \reward(i,u) - \reward(j,u')$ and select
$\Delta$ with elements $\Delta(i) = i \tilde{r}$. Clearly, $\reward_u + \Delta$ is an increasing vector, and  $\incv = (I - \discount P)^{-1} \Delta$ explicitly satisfies A1'. 

\underline{(ii) A\ref{TP2_tp},  A\ref{TP2_obs} and A\ref{obsdom}}.
A\ref{TP2_tp} and   A\ref{TP2_obs}
 are standard TP2 assumptions \cite{Lov87}; see \cite{Kri16} for several controlled sensing examples. A\ref{obsdom} is also used in \cite{Lov87}; but is not required for the controlled sensing result (statement 1 of Theorem \ref{thm:main}).

\underline{ (iii) Key new assumptions}.
     Let us focus on  A\ref{copositive}, A\ref{obssc} and A\ref{obsinit}  which are the key new assumptions that replace Assumption (c) and (f) in  \cite[Proposition 2]{Lov87}. Assumptions (c) and (f) in \cite{Lov87} are sufficient for
     $\filterd(\belief,\cdot,\action) \ls \filterd(\belief,\cdot,\action+1)$ and
     $\filter(\belief,\obs,\action) \lr \filter(\belief,\obs,\action+1)$ for all $\belief \in \Belief$. Unfortunately, Assumptions (c) and (f) in \cite{Lov87}  are mutually exclusive apart from trivial cases. 
     
     The copositive condition  A\ref{copositive} on the transition probabilities  presented in our recent
     work \cite{KP15,Kri16}   fixes  Assumption (c) in \cite{Lov87} that  $\tp(1) \leq_{\text{TP2}} \tp(2)$; such TP2 dominance only holds if $\tp(1) = \tp(2)$ or  rank 1, and so is not useful.

      Our  main new assumption is the Lehmann precision condition   A\ref{obssc} on the observation probabilities. This 
fixes  the condition (f) in  \cite{Lov87} that
      $
         \oprob_{i \obs}(2) \oprob_{i+1, \obs} \leq \oprob_{i+1,\obs}(2) \oprob_{i\obs}(1) $.
   Apart from the trivial case $\oprob(1) = \oprob(2)$, it is impossible for two stochastic matrices $\oprob(1), \oprob(2)$ to satisfy condition (f) and   A\ref{obsdom} (condition (d) in \cite{Lov87})
simultaneously.
 In  comparison,
 there is a continuum of useful examples that satisfy the conditions A\ref{obsdom} and A\ref{obssc}  (Lehmann precision)  in Theorem~\ref{thm:main}; see examples below.

 Finally, A\ref{obsinit} is an absolute continuity condition. When the observation space
 is 
 finite or has finite support, A\ref{obsinit} puts conditions on the observation probabilities at  the boundary values $\obs=1$ and $\obs=\obsdim$, and is therefore not restrictive.  A\ref{obsinit} is a sufficient condition for the range of the final
 component of the updated belief for action  $\action$ to be a subset of that
 for action $\action+1$, i.e.,
 $\{e_\statedim^\p \filter(\belief,\action,\obs), \obs \in \obspace\}  \subseteq  \{ e_\statedim^\p \filter(\belief,\action+1,\obs)$,  $\obs \in \obspace\}$.

2. {\bf Continuous observations POMDPs}: 
One specific case where A\ref{obssc} holds  is the additive noise sensing case where $\obs_k = \state_k + \snoise_k$ where  the additive  noise $\snoise_k$ is an independent and identically distributed sequence of random variables with density $p_\snoise(\cdot|\action)$. Then $\oprob_{iy} = p_\snoise(y - i|\action)$. Then it can be shown \cite{Miz06} that A\ref{obssc} holds iff $\oprob_{iy}(\action)$ is larger than $\oprob_{iy}(\action+1)$ with respect to the dispersive stochastic order.

3. {\bf Blackwell dominance vs Lehmann Precision}: As mentioned in Section \ref{sec:intro}, thus far the only known cases of structural results for controlled sensing POMDPs involves Blackwell dominance \cite{Rie91,RZ94}.
Since Theorem \ref{thm:main} uses Lehmann precision to give a  new set of conditions for controlled sensing  compared to  Blackwell dominance,
it is worthwhile comparing Blackwell dominance with Lehmann precision.

Suppose $\oprob(1) = \oprob(2) \times L$ where $L$ is a stochastic matrix. Then $\oprob(2) $ is said to Blackwell dominate $\oprob(1)$; denoted as $\oprob(2) >_{B} \oprob(1)$. Intuitively $\oprob(1)$ is noisier than $\oprob(2)$.
It is well known using a straightforward  Jensen's inequality argument that the following result holds:
\begin{theorem}[Blackwell dominance. \cite{WD80,Rie91}]
  \label{thm:blackwell}
  \begin{compactenum}
    \item  Controlled Sensing POMDP:  Suppose $\tp$ is functionally independent of the action. Then
      $\oprob(u+1) >_{B} \oprob(u)$, $u=1,\ldots,\actiondim-1$ is a  sufficient condition for the conclusion of Theorem \ref{thm:main} to hold.
    \item General POMDP: Suppose A\ref{dec_cost}, A\ref{TP2_tp},
      A\ref{TP2_obs}, A\ref{copositive} hold. Then
      $\oprob(u+1) >_{B} \oprob(u)$ is a  sufficient condition for the conclusion of Theorem \ref{thm:main} to hold.
    \end{compactenum}  
  \end{theorem}

  Blackwell dominance  exploits only the convexity of the value function.
  In comparison, Lehmann precision in Theorem~\ref{thm:main} exploits both the monotonicity and convexity of the value function. Below we discuss several examples where Blackwell dominance does not hold, but Lehmann precision holds.

  {\em  Examples}.
  (i) Here are two  examples of the observation matrices that satisfy assumptions A\ref{TP2_obs},   A\ref{obssc},  A\ref{obsinit} implying that the assumptions of  statement 1   of  Theorem \ref{thm:main} hold:  $ \statedim=3,\obsdim=3,\actiondim=2$,
  \begin{align*}
    \text{Ex1.} \quad &
                  \oprob(1) = \begin{bmatrix}
                    0.8 & 0.2 & 0 \\
                    0.1 & 0.8 & 0.1 \\
                    0 & 0.2 & 0.8 
                  \end{bmatrix}, \;
                              \oprob(2) = \begin{bmatrix}
                                0.9 & 0.1 & 0 \\ 0.2 & 0.7 & 0.1 \\ 0 & 0.2 & 0.8
                              \end{bmatrix}
\\                             
                              \text{ Ex2.} \quad  &
  \oprob(1) = \begin{bmatrix}
     0.44847   & 0.30706 &  0.24447 \\
   0.33443  &  0.28762 &   0.37795 \\
   0.32463 &   0.28971 &   0.38565 
  \end{bmatrix},
  \; \oprob(2) = \begin{bmatrix}
     0.170021  &  0.410485   & 0.419494 \\
   0.106500  &  0.433559  &  0.459941 \\
   0.020739 &   0.263223 &   0.716038
  \end{bmatrix}
  \end{align*}
Actually  for the second example above,  A\ref{obsdom} also holds implying that statement 1 and statement 2 of Theorem~\ref{thm:main} hold.
Interestingly, in both examples above, $\oprob(2)$ does not Blackwell dominate $\oprob(1)$; this illustrates the usefulness of Theorem \ref{thm:main} compared to Theorem \ref{thm:blackwell}.
\\
(ii)
Consider a controlled  sensing problem with $\statedim=\obsdim$ arbitrary positive integers, and $\actiondim =2$ sensors; choosing either sensor 1 or sensor 2
yields  a noisy  observation at most one unit different from  the Markov state, i.e.,
 $\oprob(1)$ and $\oprob(2)$  are tridiagonal matrices.
Sensor 1 is more accurate
for states $2,,\ldots,\statedim-1$, while sensor 2 is more accurate for states 1 and $\statedim$.
That is,
$\oprob_{ii}(1) = p$, $\oprob_{i,i+1}(1)=\oprob_{i,i-1}(1) = (1-p)/2$,  $\oprob_{ii}(2) = q$,
$\oprob_{i,i+1}(2) = (1-p)/2$, $\oprob_{i,i-1}(2)= (1+p)/2 - q$ with the first and last rows
as $B_{11}(1)=B_{XX}(1) = p$, $B_{12}(1) = B_{X,X-1}(1) = 1-p$, $B_{11}(2) = B_{XX}(2)> p$,
$B_{12}(2) = B_{X,X-1}(2) < 1 - p$.
Then  A\ref{TP2_obs},   A\ref{obssc},  A\ref{obsinit} hold  and so Part 1 of Theorem \ref{thm:main} holds. Blackwell dominance does not hold for this example.
\\
 (iii)  A consequence of 
\cite{Jew07} is that for  symmetric $2 \times 2$ matrices $\oprob(1), \oprob(2)$, if $\oprob_{11}(1) \leq \oprob_{11}(2)$, then Blackwell dominance is equivalent to Lehmann precision A\ref{obssc}. Also A\ref{obsinit} automatically holds. This is easy to show, see  \cite{GP10}:  $\oprob(2) >_{B} \oprob(1)$ since $L = \oprob^{-1}(2) \oprob(1)$ is a valid stochastic matrix as can be verified by explicit symbolic computation.

4. {\bf Blackwell dominance vs Lehmann Precision in Hierarchical Sensing}:  A quirk with Blackwell dominance is that the multiplication order matters.
  If  the multiplication order is reversed, i.e., suppose $\oprob(1) = M \times \oprob(2)$ where $M$ is a stochastic matrix, then even though $\oprob(1)$ is still more ``noisy'' than $\oprob(2)$, Blackwell dominance (i.e., $\oprob(1) = \oprob(2) \times L $ where $L$ is a stochastic matrix) does not necessarily hold.
%
%
%
  As an  example  consider
  $$  \statedim=3,\obsdim=3,\actiondim=2,\;
  B(1) = \begin{bmatrix}
0.3229  &  0.4703   &  0.2068\\
    	0.2237 &   0.4902  &  0.2861\\
    	0.1587  &  0.4620 &   0.3793 \end{bmatrix},\;
      B(2) = \begin{bmatrix}
0.4387   &  0.5190   &  0.0423 \\
    	0.2455  &   0.6625 &   0.0920 \\
    	0.0615   &  0.2829 &   0.6556
              \end{bmatrix}
  $$
  Then there exists a stochastic matrix $M$ such that $B(1) = M \times B(2)$ but  Blackwell dominance does not hold since  $B(1) \neq B(2)\times  L$ for stochastic matrix $L$.
  But A\ref{TP2_obs}, A\ref{obssc} (Lehmann precision) and A\ref{obsinit} hold  for this example and therefore statement 1 of Theorem \ref{thm:main} holds.

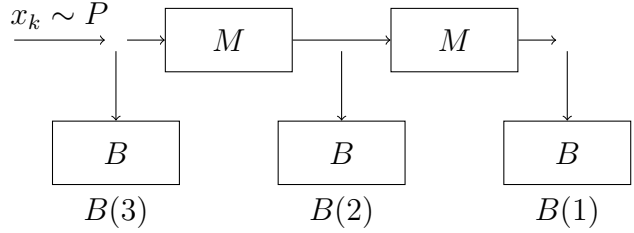
\begin{figure} \centering
\tikzstyle{block} = [draw, rectangle, minimum height=2em, minimum width=4em]
\begin{tikzpicture}[node distance = 1.5cm, auto]
  \node(block05){};
  \node[right of=block05](block15){};
   \node [block, right of=block15] (block2) {$M$};
    \node[ right of=block2] (block25) {};
    \node [block, right of=block25] (block3) {$M$};
    \node [right of=block3] (block35) {};
    \node[block,below of=block15,label=below:$\oprob(3)$](block4){$B$};
    \node[block,below of=block25,label=below:$\oprob(2)$](block5){$B$};
     \node[block,below of=block35,label=below:$\oprob(1)$](block6){$B$};
  \draw[->] (block15) -- (block2);  \draw[->] (block2) -- (block3);
    \draw[->] (block15)--(block4);
    \draw[->] (block25)--(block5);
    \draw[->](block05)-- node[pos=0.5,above]{$\state_k\sim \tp$}(block15);
    \draw[->](block3)--(block35);
     \draw[<-](block6)--(block35);
   \end{tikzpicture}
   \caption{Controlled Hierarchical Sensing where Blackwell dominance does not necessarily hold.
     Level $l$ of the network receives the Markovian signal $\state_k$ distorted by
   the confusion matrix $M^l$. Polling any specific level has observation probabilities $\oprob$; so the conditional probabilities of $\obs$ at level $l$ given $\state$ is specified by stochastic matrix $M^l \oprob$.}
   \label{fig:hierarchical}
\end{figure}

  {\em Controlled Hierarchical Sensing}.  In controlled sensing involving hierarchical sensors (such as hierarchical social networks),  level $l$ of the network
  receives signal $\state_k$ distorted  by the confusion matrix $M^l$  ($l$-th power of stochastic matrix $M$), where $l\in \{0,1,\ldots,\actiondim-1\}$.
That is,  each level of the network observes a noisy version of the previous level.
  Observing (polling) level $l$ of the network has  observation probabilities $\oprob$ conditional on the noisy message at level $l$. Therefore the conditional probabilities of the observation $\obs$ given the state $\state$ are
  $B(U-l) = M^l \times B(U)$ where $l$ is the degree of separation    from the underlying source (state).
  This is illustrated in Figure \ref{fig:hierarchical} for $\actiondim=3$.
  The controlled sensing POMDP is to choose which level to poll at each time in
  order to optimize an infinite horizon discounted reward.

  Even though
$B(u)$ is more noisy than $B(u+1)$, 
Blackwell dominance does not hold (due to the reverse multiplication order). Yet using Lehmann precision,
Theorem \ref{thm:main} holds
(under the stated assumptions).

  5. {\bf How does the optimal cumulative reward depend on Lehmann precision?} Consider two controlled sensing POMDPs with model parameters $\model=(\tp,\oprob(1),\ldots\oprob(\actiondim)$ and $\bmodel=(\tp,\bar{\oprob}(1),\ldots,\bar{\oprob}(\actiondim))$ and identical rewards.
  Let $\policy^*(\model)$ and $\policy^*(\bmodel)$ denote the corresponding optimal policies and 
let $J_{\policy^*(\model)}(\belief)$ and $J_{\policy^*(\bmodel)}(\belief)$ defined in (\ref{eq:discountedcost}) denote the respective discounted cumulative rewards when using the optimal policies.

\begin{theorem}\label{thm:ordinal} \begin{compactenum}
\item   (Lehmann precision) Suppose $\oprob(u) >_L \bar{\oprob}(u)$ for $u\in \{1,\ldots,\actiondim\}$ (see A\ref{obssc} for notation) and
  A\ref{dec_cost}, A\ref{TP2_tp}, A\ref{TP2_obs}, A\ref{obsinit}  hold. Then
  $J_{\policy^*(\model)}(\belief) \geq J_{\policy^*(\bmodel)}(\belief)$.
\item  (Blackwell dominance) Suppose $\oprob(u) >_B \bar{\oprob}(u)$ for $u\in \{1,\ldots,\actiondim\}$. Then  $J_{\policy^*(\model)}(\belief) \geq J_{\policy^*(\bmodel)}(\belief)$.
\end{compactenum}
\end{theorem}
The proof is similar to that of Statement 2 in Section \ref{sec:pmain} and thus omitted.
Even though computing the optimal policy of a POMDP is intractable,
Theorem \ref{thm:ordinal} facilitates  comparing the optimal rewards  of two different POMDPs with  different observation probabilities. Statement (2) deals with  the Blackwell dominance case; see
\cite[Theorem 14.8.1]{Kri16}. It says that in controlled sensing,  the optimal reward of a POMDP $\bmodel$ with nosier observations is smaller  than that of the POMDP $\model$; this is intuitively obvious.

Statement 1   is  more useful than Statement 2 in controlled sensing applications, since Lehmann precision does not necessarily require that  $\bmodel$ has more noisy observations than $\model$.  In controlled hierarchical sensing discussed above, Statement 1  says that 
certain networks  intrinsically yield lower optimal cumulative reward than others. For example, consider two networks where network~1 has intrinsic confusion matrix $M$ and network~2 has intrinsic confusion matrix $\bar{M} = M L$ for some stochastic matrix $L$. Then although Blackwell dominance does not hold
(due to the reverse multiplication order), Statement 1 says that controlled sensing with network 1 yields a larger cumulative reward (assuming the conditions of Theorem \ref{thm:ordinal} hold).

6.  {\bf Monotone vectors in value function for $\statedim\leq 3$}. It is well known since \cite{Son71} that the value function  $\valuef_k(\belief)= \argmax_i \gamma_i^\p \belief$ in (\ref{eq:vi}) is piecewise linear and convex in $\belief$ for any finite $k$.
Almost 40 years ago,  \cite{Alb79}  gave conditions under which the  elements of each vector $\gamma_i$ are increasing. Unfortunately the conditions in \cite{Alb79} were implicit and not easily verifiable.  As an aside,  Theorem \ref{thm:valuedec} in Sec.\ref{sec:3key} shows that under  A\ref{dec_cost},  A\ref{TP2_tp}, A\ref{TP2_obs}, Albright's result is true for $\statedim\leq 3$. 


\section{Proof of Main Result Theorem \ref{thm:main}} \label{sec:proof}
Here is some intuition. Classical convex dominance is defined for scalar convex functions $\phi: \reals\rightarrow \reals$.
In a POMDP the value function $V : \Belief \rightarrow \reals$ and so at first sight is incompatible with convex dominance.\footnote{This is why structural results which exploit convexity in POMDPs dating back to  \cite{Alb79} work with two state POMDPs.}
So the proof proceeds in two steps. First we work with the value function
on certain line segments in the unit simplex (belief space); see Figure \ref{fig:lines} for a visual illustration. On each such line segment monotone likelihood ratio dominance becomes a total order and so
the value function is convex and increasing. Because of this scalar representation of the belief  on each such line,  one can use the classical representation of the convex value function as the sum of one-dimensional wedge functions.
We then prove convex dominance of the value function in terms of such wedge functions - the key sufficient condition involves the Lehmann precision condition A\ref{obssc}.  Finally, since any belief (point) in the belief space (unit simplex) lies on one such line,  the proof holds for any belief in the simplex.

\subsection{Notation and Definitions}

{\em Monotone likelihood ratio dominance and first order dominance}
 Below $\belief(i)$ denotes the $i$-th element of belief $\belief \in \Belief$.
Let $\belief_1, \belief_2 \in \Belief$ denote  two beliefs.
$\belief_1$ dominates $\belief_2$ with respect to the MLR order, denoted as
$\belief_1 \gr \belief_2$,
 if 
 $ \belief_1(i) \belief_2(j) \leq \belief_2(i) \belief_1(j)$ $i < j$,  $i,j\in \{1,\ldots,\statedim\}$.
 $\belief_1$ dominates  $\belief_2$ with respect to first order dominance, denoted as
 $\belief_1\gs \belief_2$ if $\sum_{i\geq j} \belief_1(i) \geq \sum_{i\geq j} \belief_2(i)$ for $j \in \{1,\ldots,\statedim\}$.
 A function $\fun:\Belief\rightarrow \reals$ is said to be MLR (resp.\ first order) increasing if $\belief_1 \gr \belief_2$ (resp.\ $\belief_1 \gs \belief_2$) implies $\fun(\belief_1) \geq \fun(\belief_2)$.

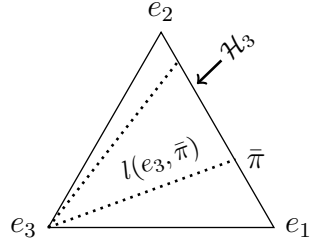
\begin{figure}\centering
\begin{tikzpicture}
  \draw [line width=0.5pt] (0,0) -- (60:3) -- (3,0) -- cycle;
  \draw [line width=1pt,dotted] (0,0) -- node[pos=0.65,above,sloped] {\footnotesize $\l(e_3,\bbelief)$} (20:2.65);
  \draw [line width=1pt,<-](44:2.75)-- node[pos=0.8,right,sloped]{\footnotesize $\Hyperplane_3$}(44:3.2);
  \draw[line width=1pt,dotted] (0,0) -- (52:2.78);
  \coordinate[label=left:$e_3$]  (C) at (0,0);
  \coordinate[label=right:$e_1$]  (A) at (3,0);
  \coordinate[label=above:$e_2$]  (B) at (60:3);
  \coordinate[label=right:$\bbelief$]  (D) at (20:2.65);
  
\end{tikzpicture}
\caption{Illustration of line segments $\l(e_\statedim,\bp)$ when $\statedim=3$. The belief space $\Pi(3)$ lies in an equilateral triangle (2-dimensional unit simplex) with vertices $e_1=[1,0,0]^\p$, $e_2=[0,1,0]^\p$ and $e_3=[0,0,1]^\p$. Any belief $\belief \in \Pi(3)$ lies on one such  dotted line $\l(e_3,\bp)$ where  belief $\bp=[\belief(1)/(1-\belief(3)),\belief(2)/(1-\belief(3)),0]^\p$ lies  on the
  hyperplane $\Hyperplane_3$ opposite $e_3$. On each line segment $\l(e_3,\bp)$ MLR dominance is a total order. Theorem  \ref{thm:valuedec}  shows that the value function is convex and increasing on each such line segment.
  Theorem \ref{thm:convexdom}  shows convex dominance on each such line segment; thereby establishing the main result Theorem \ref{thm:main}.}
\label{fig:lines}
\end{figure}

For state-space dimension $\statedim =2$, MLR is a complete order and coincides with
first order stochastic dominance.
For state-space dimension $\statedim >2$, MLR dominance implies first order dominance.
MLR is a  {\em partial order}, i.e., $[\Belief,\gr]$ is a partially ordered set (poset) since it is not always
possible to order any two belief states $\belief \in \Belief$. However, on line
segments in the simplex
defined below (see also Figure \ref{fig:lines}), MLR is a total ordering; this property is crucial for our proofs below.

Let $e_i$, $i\in \{1,2,\ldots,\statedim\}$ denote the unit $\statedim$-dimensional vector with 1 in the $i$-th position.
For $i =1$ and $i=\statedim$, define the sub simplex 
$\Hyperplane_i \subset \Belief$  as
\beq \label{eq:hi} \Hyperplane_i =  \{\belief \in \Belief:  \belief(i) = 0 \}. \eeq
Denote belief states that lie in $\Hyperplane_i$ by $\bp$.
 For each $\bp \in \Hyperplane_i$, construct the line segment $\l(e_{i},\bp)$ that connects $\bp$ to $e_{i}$. 
Thus
$\l(e_{i},\bp)$ comprises of belief states 
$\belief$ of the form:
\beq \l(e_{i},\bp) = \{\belief \in \Belief: \belief = (1-\epsilon) \bp + \epsilon e_{i}, \;
0 \leq \epsilon \leq 1 \} ,
 \bp \in \Hyperplane_i.  \label{eq:lines}
 \eeq

\begin{definition}[MLR ordering  ${\gl}$  on  lines]  \label{def:tp2l}
 $\belief_1$ is greater than $\belief_2$ with respect to the MLR ordering on
the line $\l(e_{i},\bp)$ -- denoted as $\belief_1\gl \belief_2$, if 
$\belief_1,\belief_2 \in \l(e_i,\bp)$ for  some $\bp \in \Hyperplane_i$, i.e., $\belief_1$,$\belief_2$  are on the same line connected to vertex $e_i$ of simplex $\Belief$, and
$\belief_1 \gr \belief_2$. 
\end{definition}

Note that $[\Belief,\glX]$ and $[\Belief,\glone]$ are chains\footnote{A chain is totally ordered subset of a partially ordered set.}, i.e., all elements
$\belief,\tpi \in \l(e_{X},\bp)$ are comparable, i.e., either $\belief\glX \tpi$ or $\tpi \glX \belief$
(and similarly for  $\l(e_{1},\bp)$). Figure \ref{fig:lines} illustrates this.
In Lemma \ref{lem:convex}, we summarize useful properties of $[\Belief,\gl]$ that
will be used in our proofs.

\begin{lemma} \label{lem:convex} The following properties hold on
 $[\Belief,\gr]$, $[\l(e_X,\bp),\glX]$.\\
(i) On  $[\Belief,\gr]$, $e_1$ is the least and  $e_X$ is the greatest element.
On $[\l(e_X,\bp),\gl]$, $\bp$ is the least  and $e_X$ is the greatest element.\\
(ii) Convex combinations of MLR comparable belief states form a chain. 
For any $\gamma \in [0,1]$,
$\belief \lr \tpi \implies \belief \lr \gamma\belief + (1-\gamma) \tpi \lr \tpi $.
(iii) All points on a line $\l(e_X,\bp)$ 
are MLR comparable. Consider  any two points
$\belief^{\gamma_1},\belief^{\gamma_2}\in \l(e_X,\bp)$ (\ref{eq:lines}).
Then 
$\gamma_1 \geq \gamma_2$, implies $\belief^{\gamma_1}
\gl \belief^{\gamma_2}$.  \end{lemma}

\subsection{Three key results} \label{sec:3key}
\begin{theorem}[Monotone value function] \label{thm:valuedec}
  Under A\ref{dec_cost},  A\ref{TP2_tp} and A\ref{TP2_obs}:
  \begin{compactenum}
  \item The value functions $\valuef_k(\belief)$ in (\ref{eq:vi}) and $\valuef(\belief)$ in (\ref{eq:bellman}) are MLR increasing and convex on $\Belief$.  Therefore $\valuef_k(\belief)$ and $\valuef(\belief)$ are increasing and convex on each line $\l(e_\statedim,\bp)$.
  \item (a) For any finite $k$, the value function $\valuef_k(\belief) = \max_{i \in I_k} \gamma_{i,k}^\p \belief$ 
    in (\ref{eq:vi}) is piecewise linear and convex. \\ (b) The vector $\gamma_{ik}=[\gamma_{ik}(1),\ldots,\gamma_{ik}(\statedim)]^\p$ satisfies:
$\gamma_{ik}(1) \leq \gamma_{ik}(j),j\in\{2,\ldots,\statedim-1\}  \leq \gamma_{ik}(\statedim)$. Therefore,
    for $\statedim\leq 3$,  each vector $\gamma_{ik}$ has increasing elements.
  \item On any line $\l(\bbelief,e_\statedim)$ the value function is of the form
    \beq  V_k(\belief) = \sum_{i=1}^n \max(\alpha_i e_\statedim^\p \belief - f_i,0), \quad
    \quad \belief  \in  \l(e_\statedim,\bp)
\label{eq:valueline}
    \eeq
    where $\alpha_i \geq 0$, $e_\statedim$ is the unit vector with 1 in
    the $\statedim$-th element,
and $f_i \in \reals$.
  \end{compactenum}
\end{theorem}
\proof{Proof of Theorem \ref{thm:valuedec}}
Regarding Statement 1,  \cite{Lov87} proved that the value function is MLR monotone on $\Belief$.  Convexity of the value function on the belief space 
goes back to \cite{Son71}. Therefore, the value function  is monotone and convex on each line segment $\l(e_\statedim,\bp)$.
Statement 2(a) is in  \cite{Son71}.
The proof of Statement 2(b)  follows from the fact that $V(\pi)$ is increasing
on lines towards $e_1$ which implies $\gamma_{ik}(1) \leq
\gamma_{ik}(j)$, $j=2,\ldots,\statedim$  and also  increasing on lines towards $e_X$ which implies
$\gamma_{ik}(\statedim) \geq \gamma_{ik}(j)$, $j=1,\statedim-1$.
For $\statedim=3$ this implies $\gamma_{ik}(1)\leq \gamma_{ij}(2) \leq
\gamma_{ik}(3)$.

The proof of Statement 3 is as follows: Start with Statement 2(a), namely,
$\valuef_k(\belief) = \max_{i \in I_k} \gamma_{i,k}^\p \belief$.
Obviously, all beliefs $\belief \in \Belief$ that lie on each line segment $\l(e_X,\bp)$  satisfy the straight line equation
$$ \belief = \belief(\statedim) \, e_\statedim + \big(1 - \belief(\statedim)\big) \, \bbelief , \quad \belief \in  \l(e_\statedim,\bp)$$
Therefore each piecewise linear segment $\gamma_i^\p \belief$ of the value function on the line
$\l(e_X,\bp)$  has the form
$$\gamma_i^\p \belief = 
\gamma_i^\p \bbelief + \belief(\statedim)\,  \big(\gamma_i(\statedim) - \gamma_i^\p \bbelief \big) $$
implying that for   $\belief \in  \l(e_\statedim,\bp)$, the value
function $V_k(\belief)$ has the explicit  representation
\beq  \valuef_k(\belief) = \max_{i \in I_k} \gamma_i^\p \bbelief + \belief(\statedim)\,  \big(\gamma_i(\statedim) - \gamma_i^\p \bbelief \big), 
\label{eq:explicitvalue}
\eeq
in terms of the scalar variable $\belief(\statedim) \in [0,1]$.
Statement 1 showed that $V_k(\belief)$ on each such line $\l(e_X,\bp)$ is increasing and convex.
Next, 
 any increasing convex function on
a line (i.e., a convex function that maps  $ \reals$ to $ \reals$)  is the  maximum of a countable set of increasing linear (wedge) functions; see \cite[Theorem 1.5.7]{MS02}.
Therefore, given the explicit representation (\ref{eq:explicitvalue}) of
$\valuef_k(\belief)$ in terms of the scalar variable $\belief(\statedim)$ for $\belief  \in  \l(e_\statedim,\bp)$, it follows that
for sufficiently large  $n$, $$ V_k(\belief) = \sum_{i=1}^n \max (\alpha_i \belief(\statedim) - f_i, 0),
\quad \belief(\statedim) \in [0,1],
$$
for some constants $\alpha_i \geq 0$, $f_i \in \reals$.
Equivalently,
$$ \valuef_k(\belief) = \sum_{i=1}^n \max (\alpha_i e_\statedim^\p  \belief - f_i,0) , \quad \belief  \in  \l(e_\statedim,\bp).$$
\hfill \qed

The following result is required for establishing our main result when
$\obspace$ is either finite or has finite support. A\ref{obsinit} is the crucial assumption  here.

\begin{theorem}[Finite support observation distributions]
  \label{thm:boundary} Suppose $\obspace = [a,b]$.
  Assume A\ref{TP2_tp}, A\ref{TP2_obs}, A\ref{obsinit}. Then $\{e_X^\p \filter(\belief,\obs,u), \obs \in \obspace\} \subseteq \{e_X^\p \filter(\belief,\obs,u+1), \obs \in \obspace\}$.
\end{theorem}
\proof{Proof of Theorem \ref{thm:boundary}}
Since $\filter(\belief,\obs,u) \uparrow  \obs$ under A\ref{TP2_obs} and
$\uparrow \belief$ under A\ref{TP2_tp}, it suffices to show that
\beq \label{eq:boundaryineq}
e_X^\p \filter(\belief,a,u+1) \leq e_X^\p \filter(\belief,a,u),
\quad \text{ and } \quad 
e_X^\p \filter(\belief,b,u+1) \geq e_X^\p \filter(\belief,b,u) \eeq
The first inequality in (\ref{eq:boundaryineq}) is equivalent to
$ \frac{\ones^\p \oprob_a(u) \tp^\p \belief}{\oprob_{\statedim,a}(u) e_X \tp^\p \belief}
\leq  
\frac{\ones^\p \oprob_a(u+1) \tp^\p \belief}{\oprob_{\statedim,a}(u+1) e_X \tp^\p \belief} $. Since the numerators are convex combinations
of $B_{ia}(\action)$ and  $B_{ia}(\action+1)$, $i=1,\ldots,\statedim$, respectively, A\ref{obsinit} is a sufficient condition for the inequality  to hold.
A similar proof holds for the second inequality  in (\ref{eq:boundaryineq}).

\begin{theorem}[Convex dominance for controlled sensing POMDP]
  \label{thm:convexdom}
  Suppose $\tp(\action)$ is functionally independent of $\action$.
  Assume A\ref{TP2_obs}, A\ref{obssc}, A\ref{obsinit}.  Then
  the following convex dominance holds for $\alpha> 0$:
  \begin{equation}
    \sum_{\obs \in \obspace} | \alpha e_X^\p \filter(\belief,\obs,\action)\big)- f|^+\, \filterd(\belief,\obs,\action)  \uparrow \action   \label{eq:convexdom}
  \end{equation}
\end{theorem}
\proof{Proof of Theorem \ref{thm:convexdom}}
  For notational convenience
  assume the actions are $u=1,2$. Also since\footnote{If $\alpha = 0$, the result holds trivially and there is nothing to prove.} $\alpha>0$, dividing through by $\alpha$, we need  to prove that for $\lambda \in \reals$,
\begin{equation}
  \begin{split}
    \psi(\lambda) & \ole \sum_\obs[ e_\statedim^\p \filter(\belief,\obs,2) - \lambda]^+ \filterd(\belief,\obs,2)
    - \sum_\obs [e_X^\p \filter(\belief,\obs,1) - \lambda]^+ \filterd(\belief,\obs,1)  \\
    & = \sum_{\obs \in \obspace^{\lambda}_2}  [ e_\statedim^\p \filter(\belief,\obs,2) - \lambda]\, \filterd(\belief,\obs,2) -
       \sum_{\obs \in \obspace^{\lambda}_1} [ e_\statedim^\p \filter(\belief,\obs,1) - \lambda] \, \filterd(\belief,\obs,1)
          \geq 0
         \end{split}
    \label{eq:lambdafn} \end{equation}
  where  $\obspace^\lambda_u = \{\obs: e_X^\p \filter(\belief,\obs,u) > \lambda\}$, $u=1,2$.
  Note for  $\lambda > 1$ clearly $\obspace^\lambda_u = \emptyset$ since $e_X^\p \filter(\belief,\obs,u)$ is the last component of the updated belief; and therefore $\psi(\lambda) = 0$ for $\lambda\geq 1$.
Also, for $\lambda \leq 0$, $\obspace^\lambda_u = \obspace$ and so $\psi(\lambda) = 0$ for $\lambda < 0$.
  So we only need to prove $\psi(\lambda) \geq 0$ for $\lambda \in (0,1)$.

  {\bf Case 1. $\obspace = \reals$}:
Denote $ \bar{\obspace}^\lambda_u = \obspace - \obspace^\lambda_u$ for
  $u=1,2$.
  By A\ref{TP2_obs},  $\filter(\belief,\obs,\action) \uparrow \obs$ wrt MLR order. So  $e_X^\p \filter(\belief,\obs,\action)$ is an increasing function of $\obs$.
  Define\footnote{ If $\oprob_{iy}(u)$ is discontinuous in $y$ then choose
  $y^*_{\lambda_u} = \sup\{ y:  e_X^\p \filter(\belief,\obs,\action) \leq  \lambda\}$ and assign $e_X^\p \filter(\belief, y^*_{\lambda_u},\action) = \lambda$; since $y^*_{\lambda_u}$ has measure zero it does not affect the optimal policy.}
  $y^*_{\lambda_u} = \inf \{y: e_X^\p \filter(\belief,\obs,\action) = \lambda \}$.
 Therefore 
  $\obspace^\lambda_u = ( y^*_{\lambda_u},\infty)$ for some
  $y^*_{\lambda_u}\in \reals$ and  the complement set $ \bar{\obspace}^\lambda_u = (-\infty,  y^*_{\lambda_u}]$. By absolute continuity condition A\ref{obsinit}, for $\lambda \in (0,1]$,
$\bar{\obspace}^\lambda_u $ is non-empty.
  
  We establish (\ref{eq:lambdafn})  for $\lambda \in (0,1)$ by showing\footnote{Since $\psi(0) = \psi(1) = 0$, clearly if $\psi(\lambda) \geq 0$ at its stationary points (minima), then $\psi(\lambda)\geq 0 $ for all $\lambda \in [0,1]$.} that $\psi(\lambda^*) \geq 0$
  at all stationary points $\lambda^*$ such that $d\psi(\lambda)/d\lambda = 0$.
  Note that
   \begin{equation}
    \begin{split}
      \psi(\lambda) &= \sum_{\obs \in \obspace^\lambda_2} [e_X^\p \oprob_\obs(2)\tp^\p \belief - \lambda \ones^\p  \oprob_\obs(2) \tp^\p \belief]
- \sum_{\obs \in \obspace^\lambda_1} [ e_X^\p \oprob_\obs(1)\tp^\p \belief- \lambda \ones^\p  \oprob_\obs(1) \tp^\p \belief] 
\\
&= (e_X-\lambda \ones)^\p \left[ \sum_{\obs  \in \obspace^\lambda_2}
  \oprob_\obs(2) - \sum_{\obs  \in \obspace^\lambda_1} \oprob_\obs(1) \right] \tp^\p \,\belief \\
&= (e_X-\lambda \ones)^\p \left[ \sum_{\obs  \in \bar{\obspace}^\lambda_1}
  \oprob_\obs(1) - \sum_{\obs  \in \bar{\obspace}^\lambda_2} \oprob_\obs(2) \right] \tp^\p \,\belief \\
&= \sum_{i=1}^\statedim\underbrace{ (e_X(i) - \lambda)}_{\alpha_i}\,
\underbrace{\sgn\left[ \sum_{\obs  \in \bar{\obspace}^\lambda_1}
  \oprob_{i\obs}(1) - \sum_{\obs  \in \bar{\obspace}^\lambda_2} \oprob_{i\obs}(2) \right]}_{\beta_i}\,
\underbrace{\left| \sum_{\obs  \in \bar{\obspace}^\lambda_1}
  \oprob_{i\obs}(1) - \sum_{\obs  \in \bar{\obspace}^\lambda_2} \oprob_{i\obs}(2) \right|\,
(\tp^\p \,\belief)_i}_{p_i}
\end{split} \label{eq:pre_fkg}
\end{equation}
Let us next evaluate the stationary points of  $\psi(\lambda)$ for $\lambda \in (0,1)$.

\begin{lemma} \label{lem:nabla} For  $\psi(\lambda)$ defined
  in (\ref{eq:lambdafn}), the gradient wrt $\lambda \in (0,1)$ is
\beq   \frac{d\psi(\lambda)}{d\lambda} = - \ones^{\p} \left[ \sum_{\obs  \in \bar{\obspace}^\lambda_1}
  \oprob_\obs(1) - \sum_{\obs  \in \bar{\obspace}^\lambda_2} \oprob_\obs(2) \right] \tp^\p \,\belief \eeq
\end{lemma}
(Proof at the end of this subsection).

Thus  the stationary points of $\psi(\lambda)$ satisfy
\beq \frac{d\psi(\lambda)}{d\lambda} = \ones^\p  \big[\sum_{\obs \in \bar{\obspace}^\lambda_1} \oprob_\obs(1) - \sum_{\obs \in \bar{\obspace}^\lambda_2}\oprob_\obs(2)\big] \tp^\p \belief = 
 \sum_i \beta_i p_i = 0. \label{eq:stationarypt}
\eeq
So it only remains to show that  $\psi(\lambda)$ is non-negative  at these stationary points.  To establish
 this  we use the FKG (Fortuin-Kasteleyn-Ginibre) inequality on (\ref{eq:pre_fkg}). In our framework the FKG inequality  reads: If $\alpha$, $\beta$ are generic increasing vectors
and $p$ a generic probability mass function, then
$$ \sum_i \alpha_i \beta_i p_i  \geq \sum_i \alpha_i p_i\,  \sum_i \beta_i p_i .$$
Clearly in (\ref{eq:pre_fkg}), $\alpha_i$ is increasing since the elements $(e_X-\lambda \ones)$ are increasing;  $\beta_i$ is increasing by A\ref{obssc};
$p_i$ is non-negative and thus proportional to a probability mass function.
Also from (\ref{eq:stationarypt}), $\sum_i \beta_i p_i = 0$.  So, applying  FKG inequality to
(\ref{eq:pre_fkg}) yields $\psi(\lambda) =  \sum_i \alpha_i \beta_i p_i \geq 0$.  Thus we have established~(\ref{eq:convexdom}) for $\obspace = \reals$.

{\bf Case 2. $\obspace=[a,b]$}:
Next we prove~(\ref{eq:convexdom}) for  the finite support case where 
$\obspace$ is the interval $[a,b]$.  The  key difference compared to the case $\obspace = \reals$ is that it is possible (if appropriate assumptions  are not made)  in (\ref{eq:lambdafn}) that $\obspace^\lambda_2 = \emptyset$ and $ \obspace^\lambda_1 $ is non-empty which would make $\psi(\lambda)$ defined  in
(\ref{eq:lambdafn}) negative. Assumption  
A\ref{obsinit} along with Theorem \ref{thm:boundary} prevents this from happening. Indeed, from Theorem \ref{thm:boundary},  A\ref{TP2_tp}, A\ref{TP2_obs}, A\ref{obsinit}  imply that
there are three possibilities: (i) 
  $\obspace^\lambda_2 = \emptyset$ and $ \obspace^\lambda_1 = \emptyset$: clearly  $\psi(\lambda) = 0$.
(ii)  $\obspace^\lambda_2 \neq \emptyset$ and $ \obspace^\lambda_1 = \emptyset$: clearly  from (\ref{eq:lambdafn}), $\psi(\lambda) \geq  0$.
  (iii) 
  $\obspace^\lambda_1 $ and $\obspace^\lambda_2$ are both non-empty.
  The proof for this  case follows exactly as in the proof for  $\obspace = \reals$ above. (Theorem \ref{thm:boundary} implies 
  $\obspace^\lambda_2 = \emptyset$ and $\obspace^\lambda_1 \neq \emptyset$ is impossible.)

{\bf Case 3. $\obspace$ is finite}:  Finally, we prove~(\ref{eq:convexdom}) for  the case 
  $\obspace = \{1,2,\ldots,\obsdim\}$.
  Construct  the piecewise constant probability density function
  $O_{io} = \oprob_{iy}$ for $o \in [y, y+1)$ and $y \in \{1,2,\ldots,\obsdim\}$.
It is easily seen that  $\filter(\belief,o,\action)
  = \filter(\belief,\obs,\action)$ and the value function and optimal policy remain unchanged. Then the above proof for the finite support case applies.
\hfill \qed
\\

  \proof{Proof of Lemma \ref{lem:nabla}}
  Here we prove Lemma \ref{lem:nabla} that was used to evaluate the gradient of $\psi(\lambda)$ in the proof above.
  For $t\in \reals$, define  $\obspace^t_u = \{\obs: e_X^\p \filter(\belief,\obs,u) > t\}$, $u=1,2$.
  Start with (\ref{eq:lambdafn}), and noting that
  $\sum_\obs |e_X^\p \filter(\belief,\obs,u) - \lambda|^+ \filterd(\belief,\obs,u)
  = \int_\lambda^\infty |t- \lambda|^+ \sum_y I(e_X^\p \filter(\belief,\obs,u) \geq t) dt$,
  we have
$$
\psi(\lambda) =
\int_{\lambda}^\infty |t-\lambda|^+\, \left[ \sum_{\obs \in \obspace^t_2} \filterd(\belief,\obs,2) - \sum_{\obs \in \obspace^t_1} \filterd(\belief,\obs,1)
  \right] dt
=\int_\lambda^\infty \ones^\p \left[ \sum_{\obs  \in \bar{\obspace}^t_1} B_y(1) -
  \sum_{\obs  \in \bar{\obspace}^t_2} B_y(2) \right] \tp^\p \belief\, dt $$
where the second equality follows since 
$\int_\lambda^\infty f(t) g(t) dt = f(\infty) g(\infty) - f(\lambda) g(\lambda) -
\int_\lambda^\infty g(x) d f(x) $ for generic $f,g$. Then evaluating $d\psi(\lambda)/d \lambda$ completes the proof.

A more intuitive  proof  involving Dirac delta (generalized) functions is as follows:
 From
(\ref{eq:pre_fkg}),
\begin{multline} \frac{d\psi(\lambda)}{d\lambda} =
- \ones^{\p} \left[ \sum_{\obs  \in \bar{\obspace}^\lambda_1}
  \oprob_\obs(1) - \sum_{\obs  \in \bar{\obspace}^\lambda_2} \oprob_\obs(2) \right] \tp^\p \,\belief \\ + (e_X - \lambda \ones)^\p \left[ \sum_{\obs \in \obspace}
  \delta(\lambda - e_X^\p \filter(\belief,\obs^*_{\lambda_1},1)) \,\oprob_y(1)
  - \sum_{\obs \in \obspace}
  \delta(\lambda - e_X^\p \filter(\belief,\obs^*_{\lambda_2},2))\, \oprob_y(2) \right]\,\tp^\p \,\belief 
\label{eq:stationarypt1}
\end{multline}
where $\delta(\lambda - e_X^\p \filter(\belief,\obs^*_{\lambda_u},u))$ denotes the Dirac delta function centered at
$e_X^\p \filter(\belief,\obs^*_{\lambda_u},u)$.
Next note that
$$ (e_X - \lambda \ones)^\p  \sum_{\obs \in \obspace}
\delta(\lambda - e_X^\p \filter(\belief,\obs^*_{\lambda_u},u)) \,\oprob_y(u) \, \tp^\p \belief= \big( e_X - e_X^\p \filter(\belief,\obs^*_{\lambda_u},u) \ones\big)^\p B_{\obs^*_{\lambda_u}}(u) \, \tp^\p \belief= 0$$
so that the second line of (\ref{eq:stationarypt1}) vanishes.

\subsection{Proof of Theorem \ref{thm:main}} \label{sec:pmain}
With Theorems \ref{thm:valuedec} and \ref{thm:convexdom} we can now complete the proof.\footnote{Recall A\ref{dec_cost} is not required for controlled sensing since A1' automatically holds;  we mention it here for the
general POMDP proof.}

{\em Statement 1 (Controlled Sensing)}.
Assuming A\ref{dec_cost},  A\ref{TP2_tp} and A\ref{TP2_obs},   the result (\ref{eq:valueline}) yields for all $ \belief  \in  \l(e_\statedim,\bp)$,
$$ \sum_{\obs \in \obspace} \valuef_k(\filter(\belief,\obs,\action) )\, \filterd(\belief,\obs,\action) =
\sum_{i=1}^n  \sum_{\obs \in \obspace} \max(\alpha_i e_X^\p \filter(\belief,\obs,\action)  - f_i,0) \,\filterd(\belief,\obs,\action) $$
Assuming
 A\ref{TP2_obs}, A\ref{obssc}, A\ref{obsinit}, it follows from Theorem \ref{thm:convexdom} that 
each term
$\sum_{\obs \in \obspace} \max(\alpha_i e_X^\p \filter(\belief,\obs,\action)  - f_i,0) \,\filterd(\belief,\obs,\action) \uparrow \action$. This implies
$\sum_{i=1}^n\sum_{\obs \in \obspace} \max(\alpha_i e_X^\p \filter(\belief,\obs,\action)  - f_i,0) \,\filterd(\belief,\obs,\action) \uparrow \action$.
We have thus proved that
$$ \sum_\obs \valuef_k(\filter(\belief,\obs,\action+1) )\, \filterd(\belief,\obs,\action+1) 
\geq \sum_\obs  \valuef_k(\filter(\belief,\obs,\action) )\, \filterd(\belief,\obs,\action) $$
or equivalently, in terms of the notation in (\ref{eq:vi}),
$ Q_k(\belief,u+1) - Q_k(\belief,u) \geq \reward_{u+1}^\p \belief - \reward_u^\p \belief $.
Therefore $ \reward_{u+1}^\p \belief \geq \reward_u^\p \belief \implies
\mu^*(\belief) = u+1$, i.e., $\mu_k^*(\belief) \geq \policyl_k(\belief)$ for all
$\belief \in  \l(e_\statedim,\bp)$.
Finally, any belief $\belief \in \Belief$ lies on one such  line segment $\l(e_X,\bbelief) = \{\belief: \belief = (1-\epsilon) \bbelief + \epsilon e_X\}$ where
explicitly, $\epsilon = \belief(\statedim)$ and $\bbelief(i) = \belief(i)/(1-\belief(\statedim))$, $i=1,\ldots,\statedim-1$.
Therefore, $\mu_k^*(\belief) \geq \policyl_k(\belief)$ for each
$\belief \in \Belief $.
Finally, for the infinite horizon discounted case, the value iteration algorithm (\ref{eq:vi}) converges uniformly; that is, $V_k(\belief)$ converges
uniformly to $V(\belief)$ on $\Belief$, so the results hold for $V(\belief)$.

{\em Statement 2 (General POMDP).} To simplify notation, assume $\action\in \actionspace= \{1,2\}$.
With $\valuef(\belief)$ denoting the value function of the POMDP, recall that for action $u=1$, the  POMDP parameters are $\tp(1), \oprob(1)$ and for action $u=2$, the parameters are $\tp(2),\oprob(2)$.
Define the fictitious action $u=a$ with parameters $\tp(1),\oprob(2)$.
Then Statement 1 implies that under A\ref{dec_cost}, A\ref{TP2_tp},
A\ref{TP2_obs}, A\ref{obssc}, A\ref{obsinit} that
\beq \sum_\obs \valuef(\filter(\belief,\obs,1))\, \filterd(\belief,\obs,1)
\leq \sum_\obs \valuef(\filter(\belief,\obs,a))\, \filterd(\belief,\obs,a)
\label{eq:bd1}
\eeq
since actions 2 and $a$ have the same transition matrix.
Also under copositive dominance A\ref{copositive}, $\filter(\belief,y,a)
\lr \filter(\belief,y,2)$.  From Theorem \ref{thm:valuedec}, $\valuef(\belief)$ is MLR increasing implying that $V(\filter(\belief,y,a)) \leq V(\filter(\belief,y,2))$.
Finally, A\ref{TP2_tp}-A\ref{obsdom} imply that $\filterd(\belief,\cdot,a)
\ls \filterd(\belief,\cdot,2)$. Therefore,
$$ \sum_y \valuef(\filter(\belief,\obs,a))\, \filterd(\belief,\obs,a)
\leq \sum_y \valuef(\filter(\belief,\obs,2))\, \filterd(\belief,\obs,a)
\leq \sum_y \valuef(\filter(\belief,\obs,2))\, \filterd(\belief,\obs,2)$$ Combining this with (\ref{eq:bd1}) proves the result.

\bibliographystyle{abbrv}
\bibliography{C:/Users/vikramk/styles/bib/vkm}

\end{document}